\documentclass[12pt]{article}
\usepackage{amsfonts,bm}
\usepackage{graphicx}
\baselineskip
\offinterlineskip
\begin{document}

\begin{center}
{\bf Spin Operators, Pauli Group, Commutators, Anti-Commutators,
Kronecker product and Applications}
\end{center}

\begin{center}
{\bf  Willi-Hans Steeb$^\dag$ and Yorick Hardy$^\ast$} \\[2ex]

$^\dag$
International School for Scientific Computing, \\
University of Johannesburg, Auckland Park 2006, South Africa, \\
e-mail: {\tt steebwilli@gmail.com}\\[2ex]

$^\ast$
Department of Mathematical Sciences, \\
University of South Africa, Pretoria, South Africa, \\
e-mail: {\tt hardyy@unisa.ac.za}\\[2ex]
\end{center}

\strut\hfill

{\bf Abstract.} Pauli spin matrices, Pauli group, commutators, 
anti-commutators and the Kronecker product are studied.
Applications to eigenvalue problems, exponential functions
of such matrices, spin Hamilton operators, mutually unbiased
bases, Fermi operators and Bose operators are provided.

\strut\hfill

\section{Introduction}

We investigate the commutators and anticommutators 
of Pauli spin matrices and their Kronecker product. Let
$$
\sigma_1 = \pmatrix { 0 & 1 \cr 1 & 0 }, \quad
\sigma_2 = \pmatrix { 0 & -i \cr i & 0 }, \quad
\sigma_3 = \pmatrix { 1 & 0 \cr 0 & -1 }
$$
be the Pauli spin matrices. We include $\sigma_0=I_2$,
where $I_2$ is the $2 \times 2$ unit matrix. The Pauli spin matrices
are unitary and hermitian with eigenvalues $+1$ and $-1$. 
Then the spin matrices are given by $s_1=\frac12\sigma_1$, 
$s_2=\frac12\sigma_2$, $s_3=\frac12\sigma_3$ with eigenvalues $+1/2$ and 
$-1/2$. For the Pauli spin matrices we find that 
$\sigma_1\sigma_2=i\sigma_3$, $\sigma_2\sigma_3=i\sigma_1$,
$\sigma_3\sigma_1=i\sigma_2$ and the commutators are given by 
$$
[\sigma_1,\sigma_2]=2i\sigma_3, \quad
[\sigma_2,\sigma_3]=2i\sigma_1, \quad
[\sigma_3,\sigma_1]=2i\sigma_2.
$$
The matrices $i\sigma_1$, $i\sigma_2$, $i\sigma_3$ form
a basis of the simple Lie algebra $su(2)$.  
The anti-commutators of the Pauli spin matrices vanish, i.e.
$$
[\sigma_1,\sigma_2]_+=0_2, \quad 
[\sigma_2,\sigma_3]_+=0_2, \quad 
[\sigma_3,\sigma_1]_+=0_2
$$
where $0_2$ is the $2 \times 2$ zero matrix.
Here we study the commutators and anticommutators of
the Kronecker product $\otimes$ of the Pauli spin matrices \cite{1}.
Thus we study the commutator and anticommutator of 
the $2^n \times 2^n$ unitary matrices of the form
$$
(-i)^{j_0} \bigotimes_{t=1}^n \sigma_{j_t}
$$
where $j_0\in\{0,1,2,3\}$ and $j_t\in\{0,1,2,3\}$. 
These matrices are elements of the Pauli group
\cite{2}, \cite{3}. Furthermore the square of the matrices
$$
\bigotimes_{t=1}^n \sigma_{j_t} 
$$
is the $2^n \times 2^n$ unit matrix. Then 
$$
\Pi_1 = \frac12\left(I_{2^n} + \bigotimes_{t=1}^n \sigma_{j_t}\right), \qquad
\Pi_2 = \frac12\left(I_{2^n} - \bigotimes_{t=1}^n \sigma_{j_t}\right) 
$$
are $2^n \times 2^n$ projection matrices. Whether the commutator or anticommutator
of such matrices vanishes is helpful for the eigenvalue
problem of such matrices and for the calculation 
of the exponential function of such matrices.
Other applications discussed concern spin-Hamilton operators and 
the projection to sub-Hilbert space and mutually unbiased bases. 
Finally application with Fermi operators are described.

\section{Kronecker Product of Pauli Spin Matrices}

Let us first give some examples where the commutator
or the anticommutator vanishes.
Consider first the $4 \times 4$ matrices
$$
\sigma_{12} := \sigma_1 \otimes \sigma_2, \quad
\sigma_{23} := \sigma_2 \otimes \sigma_3, \quad
\sigma_{31} := \sigma_3 \otimes \sigma_1.
$$
Then we find that the commutators vanish, i.e.
$$
[\sigma_{12},\sigma_{23}]=0_4, \quad
[\sigma_{23},\sigma_{31}]=0_4, \quad
[\sigma_{31},\sigma_{12}]=0_4
$$
where $0_4$ is the $4 \times 4$ zero matrix. For the 
anticommutators we find the hermitian invertible matrices
which can expressed as Kronecker products of the Pauli
spin matrices
\begin{eqnarray*}
[\sigma_{12},\sigma_{23}]_+ &=& 
2\pmatrix { 0 & -1 & 0 & 0 \cr -1 & 0 & 0 & 0 \cr 
           0 & 0 & 0 & 1 \cr 0 & 0 & 1 & 0 }
= -2\sigma_3 \otimes \sigma_1 
= -2\sigma_{31} \cr
[\sigma_{23},\sigma_{31}]_+ &=& 
2i\pmatrix { 0 & 0 & 0 & 1 \cr 0 & 0 & -1 & 0 \cr
             0 & 1 & 0 & 0 \cr -1 & 0 & 0 & 0 } 
= -2\sigma_1 \otimes \sigma_2 = -2\sigma_{12} \cr
[\sigma_{31},\sigma_{12}]_+ &=& 
2i\pmatrix { 0 & 0 & 1 & 0 \cr 0 & 0 & 0 & -1 \cr
            -1 & 0 & 0 & 0 \cr 0 & 1 & 0 & 0 } 
= -2\sigma_2 \otimes \sigma_3 = -2\sigma_{23}.
\end{eqnarray*}
Consider now the three hermitian and unitary $8 \times 8$ matrices
$$
\sigma_{123} := \sigma_1 \otimes \sigma_2 \otimes \sigma_3, \quad
\sigma_{312} := \sigma_3 \otimes \sigma_1 \otimes \sigma_2, \quad
\sigma_{231} := \sigma_2 \otimes \sigma_3 \otimes \sigma_1.
$$
The commutators are non-zero and we find the skew-hermitian
invertible $8 \times 8$ matrices
\begin{eqnarray*}
[\sigma_{123},\sigma_{312}] &=& 
2\pmatrix { 0_2 & 0_2 & \sigma_1 & 0_2 \cr 
            0_2 & 0_2 & 0_2 & -\sigma_1 \cr
            -\sigma_1 & 0_2 & 0_2 & 0_2 \cr 
            0_2 & \sigma_1 & 0_2 & 0_2 } 
= 2i\sigma_2 \otimes \sigma_3 \otimes \sigma_1 = 2i \sigma_{231} \cr
[\sigma_{312},\sigma_{231}] &=& 
2\pmatrix { 0_2 & 0_2 & 0_2 & \sigma_3 \cr 
            0_2 & 0_2 & -\sigma_3 & 0_2 \cr
            0_2 & \sigma_3 & 0_2 & 0_2 \cr 
            -\sigma_3 & 0_2 & 0_2 & 0_2 } 
= 2i\sigma_1 \otimes \sigma_2 \otimes \sigma_3 = 2i\sigma_{123} \cr
[\sigma_{231},\sigma_{123}] &=& 
2i\pmatrix { 0_2 & \sigma_2 & 0_2 & 0_2 \cr 
              \sigma_2 & 0_2 & 0_2 & 0_2 \cr
              0_2 & 0_2 & 0_2 & -\sigma_2 \cr 
              0_2 & 0_2 & -\sigma_2 & 0_2 }
= 2i\sigma_3 \otimes \sigma_1 \otimes \sigma_2 = 2i\sigma_{312}.
\end{eqnarray*}
However the anti-commutators vanish, i.e.
$$
[\sigma_{123},\sigma_{312}]_+ = 0_8, \quad
[\sigma_{312},\sigma_{231}]_+ = 0_8, \quad
[\sigma_{231},\sigma_{123}]_+ = 0_8.
$$
As mentioned above for the Pauli spin matrices $\sigma_1$, $\sigma_2$,
$\sigma_3$ we find that the anti-commutators vanish and
the commutators are given by $[\sigma_1,\sigma_2]=i\sigma_3$,
$[\sigma_2,\sigma_3]=i\sigma_1$, $[\sigma_3,\sigma_1]=i\sigma_2$.
Consider now the three unitary and hermitian matrices
$$
\sigma_{11} := \sigma_1 \otimes \sigma_1, \quad
\sigma_{22} := \sigma_2 \otimes \sigma_2, \quad
\sigma_{33} := \sigma_3 \otimes \sigma_3. 
$$
Then the commutators vanish, i.e. 
$$
[\sigma_{11},\sigma_{22}] = 0_4, \quad
[\sigma_{22},\sigma_{33}] = 0_4, \quad
[\sigma_{33},\sigma_{11}] = 0_4
$$
and the anticommutators can be written as Kronecker products, i.e.
\begin{eqnarray*}
[\sigma_{11},\sigma_{22}]_+ &=& -2\sigma_3 \otimes \sigma_3 = 
-2\sigma_{33} \cr
[\sigma_{22},\sigma_{33}]_+ &=& -2\sigma_1 \otimes \sigma_1 = 
-2\sigma_{11} \cr
[\sigma_{33},\sigma_{11}]_+ &=& -2\sigma_2 \otimes \sigma_2 = 
-2\sigma_{22}.
\end{eqnarray*}
Now consider the $8 \times 8$ hermitian and unitary matrices
$$
\sigma_{111} := \sigma_1 \otimes \sigma_1 \otimes \sigma_1, \quad
\sigma_{222} := \sigma_2 \otimes \sigma_2 \otimes \sigma_2, \quad
\sigma_{333} := \sigma_3 \otimes \sigma_3 \otimes \sigma_3.
$$
Here the anticommutators vanish, i.e.
$$
[\sigma_{111},\sigma_{222}]_+=0_8, \quad
[\sigma_{222},\sigma_{333}]_+=0_8, \quad
[\sigma_{333},\sigma_{111}]_+=0_8
$$
and the commutators can be written as Kronecker products, i.e.
\begin{eqnarray*}
[\sigma_{111},\sigma_{222}] &=& 
-2i\sigma_3 \otimes \sigma_3 \otimes \sigma_3 = -2i\sigma_{333} \cr
[\sigma_{222},\sigma_{333}] &=& 
-2i\sigma_1 \otimes \sigma_1 \otimes \sigma_1 = -2i\sigma_{111} \cr
[\sigma_{333},\sigma_{111}] &=&  
-2i\sigma_2 \otimes \sigma_2 \otimes \sigma_2 = -2i\sigma_{222}.
\end{eqnarray*}
Consider now the general case of the three unitary and hermitian matrices 
$$
\sigma_{11\dots 1}, \quad \sigma_{22\dots 2}, \quad 
\sigma_{33\dots 3} 
$$
with $n$ Kronecker products. If $n$ is odd the three matrices form 
a basis of a simple Lie algebra. For $n$ odd the anti-commutators vanish.
If $n$ is even the commutators vanish and the anti-commutators
can be expressed as Kronecker products of $\sigma_1$, $\sigma_2$
and $\sigma_3$.
\newline

Another useful case is that
$$
[\sigma_1 \otimes \sigma_2,\sigma_3 \otimes \sigma_3] =
[\sigma_2 \otimes \sigma_1,\sigma_3 \otimes \sigma_3] = 0_4.
$$
This implies that 
$$
[\sigma_0 \otimes \sigma_0 \otimes \sigma_1 \otimes \sigma_0 \otimes
\cdots \sigma_0 \otimes \sigma_2 \otimes \sigma_0 \otimes \cdots
\otimes \sigma_0,\sigma_3 \otimes \sigma_3 \otimes \cdots \otimes 
\sigma_3] = 0_{2^n}
$$
with $\sigma_1$ and $\sigma_2$ at the $j$'th and $k$'th position 
$(j \ne k)$ with $j,k=1,2,\dots,n$.
\newline

From the commutators given above we can also infer that the Hamilton 
operator
$$
\hat H = \frac{J_{12}}{4} 
\sum_{j=1}^n(\sigma_{1,j}\sigma_{1,j+1} + \sigma_{2,j}\sigma_{2,j+1})
+ \frac{J_3}{4}\sum_{j=1}^n(\sigma_{3,j}\sigma_{3,j+1}) 
$$ 
commutes with $\sigma_3 \otimes \sigma_3 \otimes \cdots \otimes \sigma_3$
for both both open end boundary conditions and periodic boundary 
conditions, where  
$$ 
\sigma_{\alpha,j} = \sigma_0 \otimes \cdots \sigma_0 \otimes 
\sigma_{\alpha} \otimes \sigma_0 \otimes \cdots \otimes \sigma_0
$$
with $\sigma_{\alpha}$ $(\alpha=1,2,3)$ at the $j$-th position. 

\section{Pauli Group, Commutator and Anticommutator}

The $n$-qubit Pauli group is defined by
$$
{\cal P}_n := \{ \, I_2, \, \sigma_1, \, \sigma_2, \, \sigma_3 \, \}^{\otimes n}
\otimes \{ \, \pm 1, \, \pm i \, \} 
$$
where $\sigma_1$, $\sigma_2$, $\sigma_3$ are the $2 \times 2$ Pauli
matrices and $\sigma_0 \equiv I_2$ is the $2 \times 2$ identity matrix.
The dimension of the Hilbert space under consideration is
$\dim {\cal H}=2^n$. Thus each element of the Pauli group 
${\cal P}_n$ is (up to an overall phase $\pm 1$, $\pm i$) a 
Kronecker product of Pauli matrices and $2 \times 2$ identity matrices acting on $n$ qubits. The order of the Pauli group is $2^{2n+2}$. Thus for $n=1$
we have the order 16. 
\newline

Let $j_0,\ldots, j_n,k_0,\ldots,k_n\in\{0,1,2,3\}$ and
$$ 
A = (-i)^{j_0}\bigotimes_{t=1}^n \sigma_{j_t}, \qquad
B = (-i)^{k_0}\bigotimes_{t=1}^n \sigma_{k_t}.
$$
Using the fact that
$$ 
\sigma_j\sigma_k = \delta_{j,k}I_2 + i\sum_{l=1}^3 \epsilon_{j,k,l}\sigma_l 
$$
we find, using $\epsilon_{j_t,k_t,l_t}=-\epsilon_{k_t,j_t,l_t}$,
$$
AB = (-i)^{j_0+k_0}\bigotimes_{t=1}^n
\left(\delta_{j_t,k_t}I_2+i\sum_{l_t=1}^3\epsilon_{j_t,k_t,l_t}\sigma_{l_t}\right),
$$
$$
BA = (-i)^{j_0+k_0}\bigotimes_{t=1}^n
\left(\delta_{j_t,k_t}I_2-i\sum_{l_t=1}^3\epsilon_{j_t,k_t,l_t}\sigma_{l_t}\right).
$$
Now, noting that
$$
\delta_{j_t,k_t}I_2-i\sum_{l_t=1}^3\epsilon_{j_t,k_t,l_t}\sigma_{l_t}
= (-1)^{1-\delta_{j_t,k_t}}
\left(\delta_{j_t,k_t}I_2+i\sum_{l_t=1}^3\epsilon_{j_t,k_t,l_t}\sigma_{l_t}\right),
$$
we find the following expressions for the commutator and anticommutator of $A$ and $B$
$$
[A,B] = (-i)^{j_0+k_0}\left(1-\prod_{t=1}^n(-1)^{1-\delta_{j_t,k_t}}\right)
\bigotimes_{t=1}^n
\left(\delta_{j_t,k_t}I_2+i\sum_{l_t=1}^3\epsilon_{j_t,k_t,l_t}\sigma_{l_t}\right),
$$
$$
[A,B]_+ = (-i)^{j_0+k_0}\left(1+\prod_{t=1}^n(-1)^{1-\delta_{j_t,k_t}}\right)
\bigotimes_{t=1}^n
\left(\delta_{j_t,k_t}I_2+i\sum_{l_t=1}^3\epsilon_{j_t,k_t,l_t}\sigma_{l_t}\right).
$$
Thus we obtain
$$
[A,B]=\left(1-\prod_{t=1}^n(-1)^{1-\delta_{j_t,k_t}}\right)AB,\qquad
[A,B]_+=\left(1+\prod_{t=1}^n(-1)^{1-\delta_{j_t,k_t}}\right)AB.
$$
Since $AB\neq 0$, the condition $[A,B]=0$ yields
$$
[A,B] = 0 \quad \textrm{iff}\quad \prod_{t=1}^n(-1)^{\delta_{j_t,k_t}}=(-1)^n.
$$
Consequently $[A,B]=0$ if and only if the number of coincidences
$j_t=k_t$ is even if $n$ is even, and odd when $n$ is odd.
Similarly $[A,B]_+=0$ if and only if the number of coincidences
$j_t=k_t$ is even if $n$ is odd, and odd when $n$ is even.

\section{Applications}

First we look at the eigenvalue problem. Let
$A$, $B$ be two nonzero $n \times n$ matrices.
Let $A{\bf v}=\lambda {\bf v}$ $({\bf v} \ne {\bf 0})$
be the eigenvalue equation. If $[A,B]=0_n$, then
$A(B{\bf v})=\lambda(B{\bf v})$. Consequently if $B{\bf v} \ne {\bf 0}$,
then $B{\bf v}$ is an eigenvector of the matrix $A$.
Now let us assume that the anti-commutator vanishes, i.e 
$[A,B]_+=0_n$. Then we obtain $A(B{\bf v})=-\lambda(B{\bf v})$.
Thus if $B{\bf v} \ne {\bf 0}$, then $B{\bf v}$ is an eigenvector
of $A$ corresponding to the eigenvalue $-\lambda$.
An application is given in section 2 with
$$
A = \sigma_1 \otimes \sigma_1 \otimes \sigma_1, \qquad
B = \sigma_3 \otimes \sigma_3 \otimes \sigma_3 
$$
with $[A,B]_+=0_8$. Thus since $+1$ is a eigenvalue of $A$
we find that $-1$ is an eigenvalue of $A$ with the eigenvector
$(\sigma_3 \otimes \sigma_3 \otimes \sigma_3){\bf v}$
where $A{\bf v}={\bf v}$.  
\newline

One of the main calculations in quantum theory 
in the Hilbert space ${\mathbb C}^n$ is to find 
$e^A B e^{-A}$, where $A$, $B$ are $n \times n$ matrices.
This is utilized in the solution of the Heisenberg equation of
motion. Now it is well-known that
$$
e^A B e^{-A} = B + [A,B] + \frac1{2!}[A,[A,B]] + 
\frac1{3!} [A,[A,[A,B]]] + \cdots
$$
If the commutator $[A,B]$ vanishes, we find $e^A B e^{-A}=B$.
Suppose $B$ is normal with spectral decomposition
$$
B=\sum_{j=1}^m \lambda_j\Pi_j
$$
where $\lambda_j$ are the $m$ distinct eigenvalues of $B$
and $\Pi_j$ are the projections onto the corresponding eigenspaces $V_j$.
There exists a unitary $n\times n$ matrix $U$ such that
$$
\Lambda:=UBU^*=\bigoplus_{j=1}^m\lambda_j I_{\dim(V_j)}.
$$
From $e^ABe^{-A}=B$ we obtain $\Lambda=Ue^ABe^{-A}U^*=(Ue^AU^*)\Lambda(Ue^AU^*)^{-1}$.
Since the eigenvalues $\lambda_j$ are distinct we have
$$ 
e^A=U^*\left(\bigoplus_{j=1}^m P_j\right)U 
$$
where each $P_j$ is an invertible $\dim(V_j)\times\dim(V_j)$ matrix.
When
$$ 
A = (-i)^{j_0}\bigotimes_{t=1}^n \sigma_{j_t}, \qquad
B = (-i)^{k_0}\bigotimes_{t=1}^n \sigma_{k_t}
$$
(with $A,B\neq I^{2^n}$) we find $m=2$ and $\dim(V_1)=\dim(V_2)=2^{n-1}$ so that
$A = V[I_{2^{n-1}}\oplus(-I_{2^{n-1}})]V^*$ and
$B = U[I_{2^{n-1}}\oplus(-I_{2^{n-1}})]U^*$ for some unitary $U$ and $V$.
Thus
$$
e^A = V[eI\oplus(1/e)I]V^* = U^*[P_1\oplus P_2]U.
$$
It follows that $e^{A}Be^{-A}=B$ if $(UV)[eI\oplus(1/e)I](UV)^*$ is a direct sum
of two $2^{n-1}\times 2^{n-1}$ matrices.
\newline

There is also a lesser known expansion using the 
anti-commutator (\cite{4}, \cite{5})
$$
e^A B e^A = B + [A,B]_+ + \frac1{2!} [A,[A,B]_+]_+ + 
\frac1{3!} [A,[A,[A,B]_+]_+]_+ + \cdots
$$
It follows that 
$$
e^A B e^{-A} = (B + [A,B]_+ + \frac1{2!}[A,[A,B]_+]_+ + 
\frac1{3!}[A,[A,[A,B]_+]_+]_+ + \cdots)e^{-2A}
$$
$$
e^A B e^{-A} = e^{2A}(B - [A,B]_+ + \frac1{2!}[A,[A,B]_+]_+ - 
\frac1{3!}[A,[A,[A,B]_+]_+]_+ + \cdots)
$$
If the anti-commutator of $A$ and $B$ vanishes we obtain
$$
e^A B e^{-A} = Be^{-2A} \quad \mbox{and} \quad
e^A B e^{-A} = e^{2A} B.
$$
Let 
$$
A = \sigma_{111} = \sigma_1 \otimes \sigma_1 \otimes \sigma_1, \quad 
B = \sigma_{222} = \sigma_2 \otimes \sigma_2 \otimes \sigma_2.
$$
Then $[A,B]_+=0_8$ and $A^2=I_8$ so that
$$
e^A B e^{-A} = e^{2A} B = (\cosh(2)I_8 + \sinh(2)\sigma_1 \otimes 
\sigma_1 \otimes \sigma_1)(\sigma_2 \otimes \sigma_2 \otimes \sigma_2).
$$
Another application is for spin-Hamilton operators
and projection matrices. Let $A$ be an hermitian
$d \times d$ matrix with $A^2=I_d$. Then 
$$
\Pi_+ = \frac12(I_d + A), \qquad \Pi_- = \frac12(I_d - A)
$$
are projection matrices which can be used to decompose 
the Hilbert space ${\mathbb C}^d$ into invariant sub Hilbert spaces. 
Consider for example the spin-Hamilton operators
$$
\hat H = \sum_{j=1}^2 (\sigma_j \otimes \sigma_j \otimes I_2 +
I_2 \otimes \sigma_j \otimes \sigma_j)  
$$
and 
$$
\hat K = \sum_{j=1}^2 (\sigma_j \otimes \sigma_j \otimes I_2 +
I_2 \otimes \sigma_j \otimes \sigma_j + \sigma_j \otimes I_2 \otimes 
\sigma_j). 
$$
Then both $\hat H$ and $\hat K$ commute with the operator 
$\sigma_3 \otimes \sigma_3 \otimes \sigma_3$
which is an element of the Pauli group with 
$(\sigma_3 \otimes \sigma_3 \otimes \sigma_3)^2=I_8$.
Thus we have projection matrices 
$$
\Pi_+ = \frac12(I_8 + \sigma_3 \otimes \sigma_3 \otimes \sigma_3), \quad 
\Pi_- = \frac12(I_8 - \sigma_3 \otimes \sigma_3 \otimes \sigma_3)
$$
which decomposes the Hilbert space ${\mathbb C}^8$ into
two four-dimensional sub Hilbert spaces. Then the eigenvalue
problem can be solved in these sub Hilbert spaces.
\newline

Two orthogonal bases in the Hilbert space ${\mathbb C}^d$
$$
{\cal A} = \left\{ \, {\bf e}_1, \dots, {\bf e}_d \, \right\}, \qquad
{\cal B} = \left\{ \, {\bf f}_1, \dots, {\bf f}_d \, \right\}
$$
are called unbiased if for every $1 \le j,k \le d$
(\cite{6}, \cite{7}, \cite{8}) 
$$
|\langle {\bf e}_j,{\bf f}_k \rangle| = \frac1{\sqrt{d}}.
$$
For the Pauli spin matrices $\sigma_3$, $\sigma_1$, $\sigma_2$
the normalized eigenvectors 
$$
{\cal B}_3 = \left\{ \, \pmatrix { 1 \cr 0 }, \quad \pmatrix { 0 \cr 1 } \, \right\}
$$
$$
{\cal B}_1 = \left\{ \, \frac1{\sqrt2} \pmatrix { 1 \cr 1 }, \quad 
\frac1{\sqrt2} \pmatrix { 1 \cr -1 } \, \right\}
$$  
$$
{\cal B}_2 = \left\{ \, \frac1{\sqrt2} \pmatrix { 1 \cr i }, \quad
\frac1{\sqrt2} \pmatrix { 1 \cr -i } \, \right\}
$$
each form an orthonormal basis in ${\mathbb C}^2$. 
Furthermore this is a set of mutually unbiased bases. 
Consider now $\sigma_3 \otimes \sigma_3$, $\sigma_1 \otimes \sigma_1$,
$\sigma_2 \otimes \sigma_2$. Then
$$
{\cal B}_3 = \left\{ \, \pmatrix { 1 \cr 0 } \otimes \pmatrix { 1 \cr 0 }, \,\, 
\pmatrix { 1 \cr 0 } \otimes \pmatrix { 0 \cr 1 }, \,\,
\pmatrix { 0 \cr 1 } \otimes \pmatrix { 1 \cr 0 }, \,\,
\pmatrix { 0 \cr 1 } \otimes \pmatrix { 0 \cr 1 } \, \right\}
$$
$$
{\cal B}_1 = \left\{ \, \frac12 \pmatrix { 1 \cr 1 } \otimes \pmatrix { 1 \cr 1 },
\,\,
\frac12 \pmatrix { 1 \cr 1 } \otimes \pmatrix { 1 \cr -1 }, \,\,
\frac12 \pmatrix { 1 \cr -1 } \otimes \pmatrix { 1 \cr 1 }, \,\,
\frac12 \pmatrix { 1 \cr -1 } \otimes \pmatrix { 1 \cr -1 } \, \right\}
$$  
$$
{\cal B}_2 = \left\{ \, \frac12 \pmatrix { 1 \cr i } \otimes \pmatrix { 1 \cr i },
\,\,
\frac12 \pmatrix { 1 \cr i } \otimes \pmatrix { 1 \cr -i }, \,\,
\frac12 \pmatrix { 1 \cr -i } \otimes \pmatrix { 1 \cr i }, \,\,
\frac12 \pmatrix { 1 \cr -i } \otimes \pmatrix { 1 \cr -i } \, \right\}
$$
provide mutually unbiased bases in ${\mathbb C}^4$.
\newline

Another application is with Fermi operators. Let $c_1^\dagger$, $c_2^\dagger$, $c_1$, $c_2$ be Fermi 
creation and annihilation operators with the anticommutation relations
$$
[c_j,c^\dagger_k]_+ = \delta_{jk}I, \quad j,k=1,2 
$$
and $[c_j,c_k]_+=[c_j^\dagger,c_k^\dagger]_+=0$ for $j,k=1,2$.
Using the Pauli spin matrices we can form the operators
\begin{eqnarray*}
X_1 = \pmatrix { c_1^\dagger & c_2^\dagger }
\sigma_1 \pmatrix { c_1 \cr c_2 } &=&
c_1^\dagger c_2 + c_2^\dagger c_1 \\ 
X_2 = \pmatrix { c_1^\dagger & c_2^\dagger }
\sigma_2 \pmatrix { c_1 \cr c_2 } &=&
-ic_1^\dagger c_2 + ic_2^\dagger c_1 \\
X_3 = \pmatrix { c_1^\dagger & c_2^\dagger }
\sigma_3 \pmatrix { c_1 \cr c_2 } &=&
c_1^\dagger c_1 - c_2^\dagger c_2.
\end{eqnarray*}
Then we find the well-known result for the commutators
$$
[X_1,X_2] = 2iX_3, \quad [X_2,X_3] = 2iX_1, \quad 
[X_3,X_1] = 2iX_2.
$$
For the anti-commutators we find 
$$
[X_1,X_2]_+ = 0, \quad [X_2,X_3]_+ = 0, \quad [X_3,X_1]_+ = 0.
$$
This can be extended to higher dimensions. Considering Bose
creation and annihilation operators $b_1^\dagger$, $b_2^\dagger$,
$b_1$, $b_2$ and 
\begin{eqnarray*}
Y_1 = \pmatrix { b_1^\dagger & b_2^\dagger }
\sigma_1 \pmatrix { b_1 \cr b_2 } &=&
b_1^\dagger b_2 + b_2^\dagger b_1 \\ 
Y_2 = \pmatrix { b_1^\dagger & b_2^\dagger }
\sigma_2 \pmatrix { b_1 \cr b_2 } &=&
-ib_1^\dagger b_2 + ib_2^\dagger b_1 \\
Y_3 = \pmatrix { b_1^\dagger & b_2^\dagger }
\sigma_3 \pmatrix { b_1 \cr b_2 } &=&
b_1^\dagger b_1 - b_2^\dagger b_2
\end{eqnarray*}
with the commutation relations $[b_j,b_k^\dagger]=\delta_{jk}I$
$(j,k=1,2)$ provides the well-known result
$$
[Y_1,Y_2] = 2iY_3, \quad [Y_2,Y_3] = 2iY_1, \quad [Y_3,Y_1] = 2iY_2
$$
for the commutators.

\section{Conclusion}

We studied the Pauli group and found the conditions when 
the commutator or anticommutator of two elements  
of the Pauli group vanish. Six applications 
for these conditions have been discussed.
\newline

{\bf Acknowledgment}
\newline

The authors are supported by the National Research Foundation (NRF),
South Africa. This work is based upon research supported by the National
Research Foundation. Any opinion, findings and conclusions or recommendations
expressed in this material are those of the author(s) and therefore the
NRF do not accept any liability in regard thereto.

\end{document}